\RequirePackage{fix-cm}
\documentclass[twocolumn]{svjour3}          
\smartqed  
\usepackage{graphicx}
\usepackage{subfigure}

\begin{document}

\title{Current-voltage characteristics and vortex dynamics in highly underdoped La$_{2-x}$Sr$_{x}$CuO$_{4}$
}

\author{Zhenzhong Shi         \and
        P. G. Baity \and
        Dragana Popovi\'{c} 
}

\institute{Zhenzhong Shi \and P. G. Baity \and Dragana Popovi\'{c}$^\ddag$\at
              National High Magnetic Field Laboratory,Florida State University, 1800 E. Paul Dirac Drive, Talahassee, FL 32310 USA; \\             
              \ddag\email{dragana@magnet.fsu.edu}  
}

\date{Received: date / Accepted: date}

\maketitle


\begin{abstract}
The temperature dependence of the nonlinear current-voltage ($I$-$V$) characteristics in highly underdoped La$_{2-x}$Sr$_{x}$CuO$_{4}$ (x=0.07 and 0.08)
thick films has been studied in both zero and perpendicular magnetic fields $H$.  Power-law behavior of $V(I)$ is found for both $H=0$ and $H \neq 0$.  The critical current $I_{c}$ was extracted, and its temperature and magnetic field dependences were studied in detail.   The Berezinskii-Kosterlitz-Thouless physics dominates the nonlinear $I$-$V$ near the superconducting transition at $H=0$, and it continues to contribute up to a characteristic temperature $T_x(H)$.  Nonlinear $I$-$V$ persists up to an even higher temperature $T_{h}(H)$ due to the depinning of vortices.

\end{abstract}


\section{Introduction}
\label{intro}

In underdoped cuprates, which are layered materials with weak interlayer coupling, studies of paraconductivity have shown a strong 2D character of the superconducting fluctuations \cite{Leridon2007,RullierAlbenque2011}. However, the existence of a Berezinskii-Kosterlitz-Thouless (BKT) transition in bulk cuprates has been controversial \cite{Emery1995,Benfatto2007,Baity2015}. For example, the penetration depth measurements in bulk YBa$_{2}$Cu$_{3}$O$_{7-x}$ found no signatures of the BKT physics in thick films \cite{Broun2007} or crystals \cite{Kamal1994}, while dc transport measurements suggested BKT-like behavior 
in bulk samples of several cuprates (see, \textit{e.g.}, Refs. \cite{forro_prb88,yeh_prb89,norton_prb93,coton2013}).  A recent study of the paraconductivity and the $I$-$V$ characteristics in highly underdoped La$_{2-x}$Sr$_{x}$CuO$_{4}$ ($x=0.07$ and $ 0.08$) thick films (150 \AA\, CuO$_2$ layers)
in zero magnetic field \cite{Baity2015} showed that the effective dimensionality of the samples is 2D, and that the thermally-driven transition to the superconducting state is of the BKT type with a large vortex-core energy $\mu\approx 1.4 \mu_{XY}$ ($\mu_{XY}=\pi^{2}J_s/2$ is the conventional value that it assumes in the XY model; $J_s$ is the superfluid stiffness).  At the BKT transition, $J_{s}$ jumps from 0 above $T_{BKT}$ to $2{T_{BKT}}/{\pi}$ at $T_{BKT}^{-}$, leading to a change from linear to super-linear behavior in the $I$-$V$ characteristics: $V{\propto}I^{a(T)}$, where ${a(T)}$ ($={{\pi}{J_{s}}/T}+1$) jumps from 1 to 3. $T_{BKT}$ 
was determined to be 4.0~K and 9.7~K for the $x=0.07$ and $x=0.08$ samples, respectively \cite{Baity2015}. The presence of inhomogeneities leads to some smearing of the transition, giving rise to finite superfluid stiffness even for $T>T_{BKT}$ \cite{Baity2015}.  Here we present a detailed study of the evolution of the $I-V$ characteristics of the same La$_{2-x}$Sr$_{x}$CuO$_{4}$  (LSCO) samples in a perpendicular magnetic field $H$.

At small fields, $H$-induced free vortices coexist with BKT-like thermally generated vortex-antivortex pairs \cite{Minnhagen1981,Benfatto2012,Li2005a}. As $H$ increases, vortex-antivortex pairs break and $H$-induced free vortices proliferate, leading to novel phases of the vortex matter \cite{Blatter1994,LeDoussal2010}. Indeed, in the same underdoped LSCO films, two quantum critical points have been found to associate with boundaries of different vortex phases \cite{Shi2014}. 
In general, both the BKT physics and the physics of vortex matter have been subjects of intensive studies, but they have been mostly treated as two separate topics \cite{Benfatto2007,Blatter1994,LeDoussal2010}. Some efforts have been put forward to treat the BKT physics in a magnetic field and vortex matter physics on the same footing \cite{Tesanovic1999,Gasparov2012}, though much remains to be understood. The goal of our study is to bridge the BKT physics and the physics of vortex matter. 

\section{Experiment}
\label{Exp}

The samples were $\approx$100~nm thick LSCO films with $x = 0.07$ and $x = 0.08$, grown by molecular beam epitaxy.  The samples have been described in detail elsewhere \cite{Baity2015,Shi2013}. The mean-field transition temperature $T_{c}$ was determined to be 6.5~K and 11.3~K for the $x = 0.07$ and $x = 0.08$ samples, respectively \cite{Baity2015}. Standard four-probe $I$-$V$ measurements were performed up to 9 T ($H$ ${\parallel}$ c-axis). A low-pass pi-filter was used to eliminate current noise that leads to Ohmic tails at low excitations \cite{Sullivan2004}. Great care was taken to exclude Joule heating effects, including (1) a comparison of the $I$-$V$ measurements using dc current and pulsed current (50 ${\mu}$s pulse width) and (2) a comparison of the $R$ vs. $I$ curve with the $R$ vs. $T$ curve for different $H$. 

From the $I$-$V$ characteristics, the critical current $I_{c}$ is determined as the intersection of a power-law fit at 100~nV, just above the noise background. At large $H$ where BKT physics is absent, this $I_{c}$ indicates the onset of depinning in the vortex matter description, similar to previous studies in conventional superconductors \cite{Duarte1996} and in cuprates \cite{Safar1995}. For the zero-field BKT transition, however, the regime for non-linear voltage response is bounded by lower and upper critical currents. Below the lower critical current, the system exhibits Ohmic behavior due to finite-size effects \cite{Pierson1999}. On the other hand, above the upper critical current, Ohmic behavior occurs from Cooper pair depairing and a return to the normal state \cite{Reyren2007}. Although our definition of $I_c$ is, strictly speaking, neither of these, $I_{c}$ as defined here will still be a measure of excitation needed to observe nonlinear $I$-$V$ in the sample. Most importantly, since similar definitions of $I_c$ have already been used in other studies of $I$-$V$ characteristics in both $H=0$ \cite{Minnhagen1992} and $H{\neq}0$ cases \cite{Duarte1996,Safar1995}, using this criterion for $I_{c}$ becomes a bridge that helps address the crossover regime at small fields where both BKT physics and vortex matter dynamics are relevant.

\section{Results and Discussion}
\label{Results}

\begin{figure}
\includegraphics[width=0.47\textwidth]{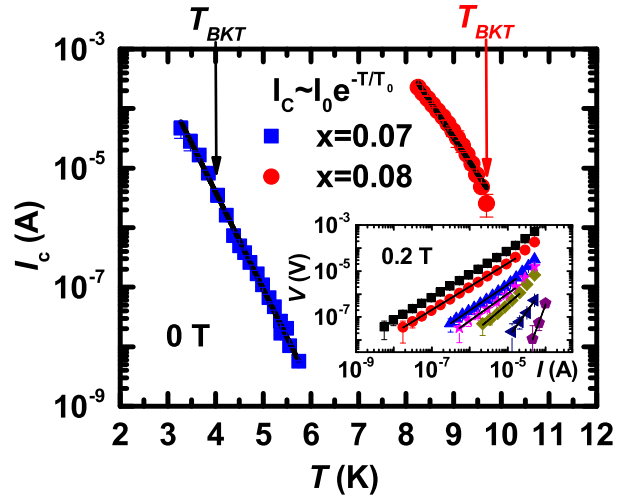}\label{Fig1}
\caption{$I_{c}$ vs. $T$ is plotted on a semi-log scale for the $x=0.07$ and $x=0.08$ samples at $H$ = 0. $T_{BKT}$ is marked with arrows. Dashed lines show exponential fits $I_{c}={I_{0}}e^{-T/T_{0}}$ for both samples, where $T_{0}$ is $0.27{\pm}0.01$ K and $0.40{\pm}0.01$ K for $x=0.07$ and $x = 0.08$, respectively.  Inset:  $V(I)$ for the x=0.07 sample in $H=0.2$~T, plotted on a log-log scale for $T$ = 4.5 K, 4.0 K, 3.4 K, 3.2 K, 3.0 K, 2.5 K and 2.0 K (from top to bottom). Solid lines are fits.}
\label{IV_0T}
\end{figure}

From the $I$-$V$ measurements at $H$ = 0 for the $x = 0.07$ and $x = 0.08$ samples, $I_{c}(T)$ were extracted, as shown in Fig.~\ref{IV_0T}. Near $T_{BKT}$, $I_{c}$ decreases exponentially as $T$ increases and disappears when the superfluid stiffness $J_s$ becomes zero and the exponent $a$ becomes 1 (\textit{e.g.} near 6~K for the $x=0.07$ sample \cite{Baity2015}). The data are best fitted with $I_{c}={I_{0}}e^{-T/T_{0}}$, where $T_{0}$ is $0.27{\pm}0.01$~K and $0.35{\pm}0.01$~K for the $x = 0.07$ and $x = 0.08$ samples, respectively. The strong, exponential dependence of $I_{c}(T)$ at $H=0$ is an interesting result and has not been reported to the best of our knowledge.

The nonlinear $V(I)$ characteristics in $H{\neq}0$ have been studied in detail for the $x = 0.07$ sample. Figure~\ref{IV_0T} inset shows the typical traces of $V(I)$ at 0.2~T for various $T$ from 2.0~K to 4.5~K. Power-law fits $V{\propto}I^{a(T)}$ were obtained in the lowest current regime at each $T$, and a change from nonlinear $V(I)$ at low $T$ to linear $V(I)$ at high $T$ was observed. This power-law behavior of $V(I)$ at $H \neq 0$ bears a close resemblance to that at $H=0$, though the underlying physics is more complicated. First, it could have contributions from the BKT physics\cite{Minnhagen1981,Benfatto2012,Li2005a,Tesanovic1999}. In addition, the presence of a finite $H$ facilitates the formation of the vortices along the field direction and suppresses the vortices in the opposite direction. As $H$ is increased, the density of vortices increases and the BKT physics becomes less relevant \cite{LeDoussal2010}. Vortices are pinned as soon as they are formed and a critical force (current) is needed to depin them. Therefore, the vortex matter dynamics also contributes to the power-law behavior of $V(I)$. The role of excitation current changes from breaking apart vortex-antivortex
pairs in the BKT scenario to inducing flux creeps (for small $I$) and depinning the vortices (for large $I$) in the vortex matter scenario. We note that we did not observe any exponential $V(I)$, such as that suggested by Kim-Anderson flux creep model \cite{ANDERSON1964}. 


\begin{figure}
\includegraphics[width=0.45\textwidth]{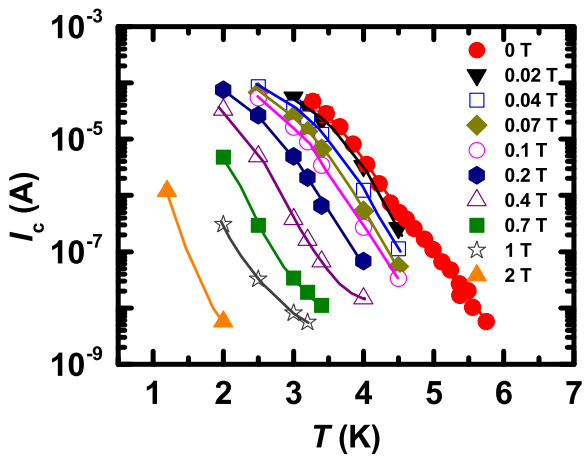}\label{Fig2}
\caption{Critical current $I_{c}$ vs. $T$ for several $H$ for the $x=0.07$ sample. The solid lines guide the eye.}
\label{IV_T}
\end{figure}


As $H$ increases, $I_{c}$ at a given $T$ is suppressed, but $I_{c}(T)$ remains near-exponential (Fig.~\ref{IV_T}). The fact that $I_{c}(T)$ does not change abruptly seems to suggest a cross-over regime in the $H-T$ phase diagram where the BKT physics is still important, consistent with magnetometry measurements in Bi$_{2}$Sr$_{2}$CaCu$_{2}$O$_{8}$ single crystals \cite{Li2005a}. Above $\sim$ 0.4~T, the curvature of the $I_{c}(T)$ on the semi-log scale changes its sign, and $I_{c}(T)$ no longer resembles its zero-field counterpart, suggesting that the effect of BKT physics becomes negligible.  Different regimes can be distinguished more clearly from the $I_c(H)$ dependence, as discussed below.  Here we note that the observed near-exponential decrease of $I_{c}(T)$ is consistent with early studies on vortex matter in other cuprates at much higher doping \cite{Safar1995,Senoussi1988}. It was attributed to the melting of the vortex lattice \cite{Safar1995} or a vortex glass phase with ${I_{c}}{\propto}{I_{0}}$exp$[-({T/{T_{0}}})^{n}]$, where $n$ depends on the ratio of the electronic mean free path to the superconducting coherence length \cite{Senoussi1988}. Here we suggest that the BKT physics may also need to be taken into account.

\begin{figure}
\includegraphics[width=0.45\textwidth]{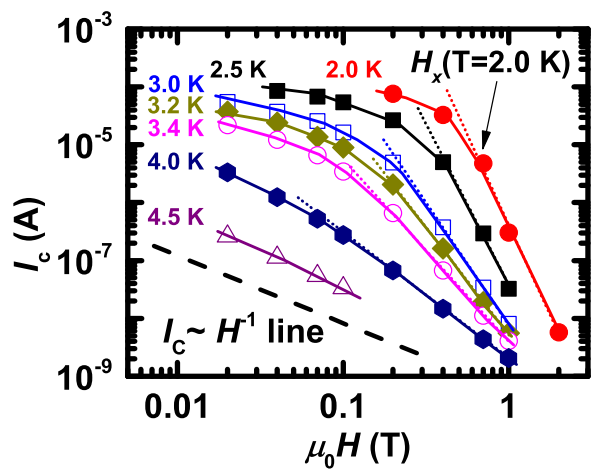}\label{Fig3}
\caption{Critical current $I_{c}$ vs. $H$ at several $T$ for the $x=0.07$ sample. Solid lines guide the eye.  The dashed line represents $I_{c}{\propto}H^{-1}$, which is expected from the Larkin-Ovchinnikov theory of collective pinning \cite{Larkin1979}.  Dotted lines are linear fits, on a log-log scale, to the data in the high-$H$ regime.  The crossover fields $H_x(T)$ are identified as fields below which the $I_c(H)$ data start to deviate from the high-field fits.}
\label{IV_H}
\end{figure}

Figure~\ref{IV_H} shows that $I_{c}(H)$ at low $T$ exhibits a kink-like feature that separates the low-$H$ regime, where $I_{c}$ varies relatively slowly with $H$, from the high-$H$ regime, where $I_{c}$ drops sharply with $H$. This kink-like feature, as well as the nonlinearity of $V(I)$, are suppressed with increasing $T$, and disappear above 4.5~K.  Two different regimes of $I_{c}(H)$ were also observed in some previous studies of conventional superconductors \cite{Duarte1996,zhu2010,Petrovic2009}. In those studies, $I_{c}$ is independent of $H$ in the low-$H$ regime, and it evolves into a power-law behavior, $I_{c}{\propto}H^{m}$ with $m\sim-1$, in the high-$H$ regime. These observations were interpreted as signatures of individual vortex pinning in the low-$H$ regime and collective pinning in the high-$H$ regime. In Fig.~\ref{IV_H}, a $I_{c}{\propto}H^{-1}$ line, expected from the Larkin-Ovchinnikov (LO) theory of collective pinning \cite{Larkin1979}, is drawn for comparison.  In our samples, $I_{c}$ does tend to saturate at the lowest $H$, consistent with individual vortex pinning, but it decreases much faster in the high-$H$ regime compared to the prediction of the LO theory (\textit{e.g.} $m\sim -6$ at 2.0~K). The rather sharp drop of $I_{c}$ at high $H$ indicates that the weak collective pinning picture of the LO theory is not sufficient in our highly underdoped LSCO films.

\begin{figure}
\includegraphics[width=0.45\textwidth]{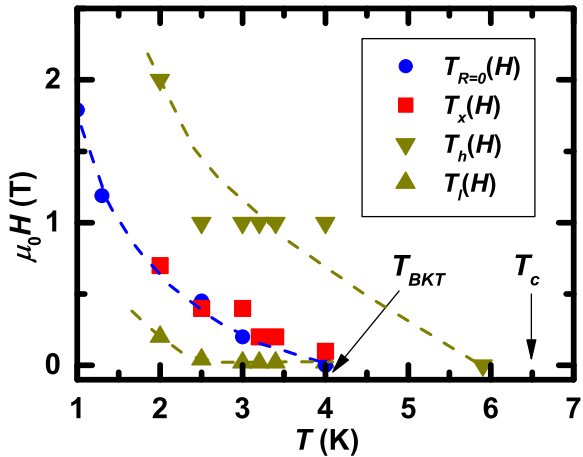}\label{Fig4}
\caption{The $H-T$ phase diagram determined from the $I-V$ characteristics for the $x=0.07$ sample.  $T_x(H)$ denotes the crossover between the low-$H$ and high-$H$ regimes in Fig.~\ref{IV_H}.  $T_{R=0}(H)$ is the temperature at which the resistance goes to zero \cite{Shi2014}.  $T_{h}(H)$ is the temperature scale above which $V(I)$ becomes linear. $T_{l}(H)$ marks the temperature below which there are no data available. $T_{BKT}= 4.0$~K and $T_{c}= 6.5$~K \cite{Baity2015} are marked by arrows.  Dashed lines guide the eye.}
\label{IV_sum}
\end{figure}

The onset of the high-$H$ regime is identified as shown in Fig.~\ref{IV_H}.  It is striking that the corresponding crossover fields $H_x(T)$, \textit{i.e.} crossover temperatures $T_x(H)$, follow closely $T_{R=0}(H)$ (Fig.~\ref{IV_sum}), the temperatures at which the resistance drops to zero.  The $T_{R=0}(H)$ line was interpreted as the transition between a pinned vortex solid (Bragg glass) and an unpinned one \cite{Shi2014}.  Nevertheless, $V(I)$ remains nonlinear up to $T_{h}(H)$ (the discrete drop and the apparent step in $T_{h}(H)$ in Fig.~\ref{IV_sum} are due to the limited resolution of our data; we expect $T_{h}(H)$ to change smoothly as suggested by the dashed guide line).  $T_{l}(H)$ in Fig.~\ref{IV_sum} is the temperature below which there are no data available. Therefore, we identify three regimes in the phase diagram.  Up to $T_x(H)$, $V(I)$ is nonlinear due to both the BKT physics and the depinning of vortices. Between $T_x(H)$ and $T_{h}(H)$, the nonlinear $V(I)$ is due to the depinning of vortices and the BKT physics is not relevant. Above $T_{h}(H)$, $V(I)$ is linear and vortices are no longer pinned.  

Finally, since the presence of inhomogeneities leads to finite $J_{s}$ up to $\sim 6$~K ($>T_{BKT}=4$~K) in $H=0$ \cite{Baity2015}, we note that it is reasonable to expect that their effect may be important up to $T_h(H)$, giving rise to nonlinear $V(I)$ over extended $T$ and $H$ ranges.  Interestingly, signatures of the BKT physics in a magnetic field have also been observed in strongly disordered conventional superconductor films \cite{Misra2013}. It was suggested that the pinning and creep of vortices in a strongly disordered vortex lattice dominates over melting phenomena associated with a clean system.  Therefore, it would be interesting to extend our studies to LSCO at higher doping where the inhomogeneity is smaller.

\section{Summary}
\label{Sum}

Our study of the nonlinear $V(I)$ in highly underdoped thick films of La$_{2-x}$Sr$_{x}$CuO$_{4}$ at different $T$ and perpendicular $H$ reveals a change in the vortex dynamics at a characteristic temperature $T_x(H)$.  In $H=0$, the superconducting transition is of the BKT type \cite{Baity2015}. As $H$ increases, the nonlinear $V(I)$ has its origin in both the BKT physics and the depinning of vortices until $T_x(H)$ is reached. Above $T_x(H)$, the BKT physics becomes irrelevant but the depinning of vortices continues to contribute to nonlinear $V(I)$ up to an even higher temperature $T_{h}(H)$. 


\begin{acknowledgements}
We thank A. T. Bollinger and I. Bo\v{z}ovi\'{c} for the samples. We acknowledge L. Benfatto for useful discussions. This work was partially supported by NSF Grant No. DMR-1307075 and by the NHMFL, which is supported by NSF/DMR-1157490 and the State of Florida.
\end{acknowledgements}


\end{document}